\documentclass{ifacconf}

\usepackage{graphicx, color}      % include this line if your document contains figures
\usepackage{natbib}        % required for bibliography
\usepackage{caption}
\usepackage{algorithm}
\usepackage{epsfig}
\usepackage{psfrag}

\usepackage{fancyhdr}

\usepackage{amsmath} % assumes amsmath package installed
\usepackage{amssymb}  % assumes amsmath package installed
\usepackage{dsfont}  % assumes amsmath package installed
\usepackage{todonotes}
\usepackage{tikz}
\usetikzlibrary{arrows,shapes,automata,backgrounds,petri,calc}

\usepackage{booktabs}
\usepackage[export]{adjustbox}

\usepackage{caption}
\usepackage{subcaption}

\newtheorem{remark}{Remark}

\DeclareMathOperator{\E}{\mathsf{E}}

\begin{document}
\begin{frontmatter}

\title{Event-triggered Add-on Safety for Connected and Automated Vehicles Using Road-side Network Infrastructure} 

%\thanks[footnoteinfo]{}

\author[First]{Mohammad H. Mamduhi} 
\author[Second]{Ehsan Hashemi} 
\author[First,Third]{John S. Baras}
\author[First]{Karl H. Johansson}

\address[First]{Division of Decision and Control Systems, School of Electrical Engineering and Computer Science, Royal Institute of Technology, Stockholm 10044, Sweden (e-mail: \{mamduhi,baras,kallej\}@ kth.se).}
\address[Second]{Mechanical and Mechatronics Engineering Department, University of Waterloo, ON, Canada (e-mail: ehashemi@uwaterloo.ca)}
\address[Third]{Institute of Systems Research, Department of Electrical and Computer Engineering, University of Maryland College Park, MD 20742, USA (e-mail: baras@umd.edu)}

\begin{abstract}                % Abstract of not more than 250 words.
%The vehicles maneuverability is significantly affected by the combined-slip friction effect, in which larger longitudinal tire slips result in considerable drop in lateral tire forces. This is of higher importance when unexpected dangerous situations occur on the road and immediate actions, such as braking, are taken to avoid collision. Harsh barking can lead to high-slip and possible loss of maneuverability, hence timely braking is essential in order to reduce high-slip scenarios while avoiding collision. In addition to the vehicle's active-safety mechanism, we propose a danger-triggered add-on safety mechanism that is activated upon being informed about danger stimuli by the road-side infrastructure. The aim is to incorporate the add-on safety mechanism to adjust the control parameters for timely and smooth braking such that a collision is avoided while vehicle's maneuverability is maintained. We study two different wireless technologies for communication between the vehicles and base stations, the Long-Term Evolution (LTE) and the fifth generation (5G) schemes. We evaluate the advantages of including the add-on safety mechanism to augment the safety margins for each communication scenario and compare the results without the proposed safety mechanism. The proposed framework is validated by utilizing CarSim that is a high-fidelity software for vehicle simulations. 

This paper proposes an event-triggered add-on safety mechanism to adjust the control parameters for timely braking in a networked vehicular system while maintaining maneuverability. Passenger vehicle maneuverability is significantly affected by the combined-slip friction effect, in which larger longitudinal tire slips result in considerable drop in lateral tire forces. This is of higher importance when unexpected dangerous situations occur on the road and immediate actions, such as braking, need to be taken to avoid collision. Harsh braking can lead to high-slip and loss of maneuverability; hence, timely braking is essential to reduce high-slip scenarios. In addition to the vehicles own active safety systems, the proposed event-triggered add-on safety is activated upon being informed about dangers by the road-side infrastructure. The aim is to incorporate the add-on safety feature to adjust the automatic control parameters for smooth and timely braking such that a collision is avoided while vehicle's maneuverability is maintained. We study two different wireless technologies for communication between the infrastructure and the vehicles, the Long-Term Evolution (LTE) and the fifth generation (5G) schemes. The framework is validated through high-fidelity software simulations and the advantages of including the add-on feature to augment the safety margins for each communication technology is evaluated.
\end{abstract}


\end{frontmatter}
%===============================================================================

\section{Introduction}
The potential to enable fast and reconfigurable mechanisms for increasingly prevalent Automated Driving Systems (ADS), cooperative vehicles, and advanced driver-assistance systems (ADAS) underscores the critical need to develop more reliable safety mechanisms within the distributed-system framework for intelligent transportation systems. The functionality of mentioned mechanisms has gradually shifted from stabilization of vehicle/wheel dynamics to taking control in emergency cases and guidance control (\cite{bengler2014three, li2017three, linsenmayer2017event}).  Connected vehicles with partially or conditionally automated driving features (known as Level 3 to Level 4 in vehicles autonomy)  not only utilize several ADS control systems, such as differential braking, torque vectoring, and active steering, but also can benefit from shared information over network nodes and infrastructure to have more proactive and reliable lateral/longitudinal stability control and to enable dramatic improvements in fuel efficiency, see \cite{lu2014connected, chang2015estimated, hung2019vehicle}. Differential braking strategies in vehicle active safety systems are capable of stabilizing the vehicle not only by expanding the safe operating envelope due to optimal speed reduction, but through generating corrective yaw moment to enhance yaw tracking performance based on the request from the path planner or the driver. Having information about road emergency cases significantly improves performance of the braking actuation due to the fact that local vehicle controllers apply corrective longitudinal forces less aggressively, leading to less longitudinal slip (\cite{ito2018coordination}). This results in less lateral tire force drop, thus, increases maneuverability and capability of the automated driving system (or the driver) in obstacle avoidance, sudden path changes, and speed reduction.  

New communication technologies such as 5G, provide fast, flexible and application-oriented network services that can be adjusted according to user-demands or criticality of the situation, \cite{MOLINA2018407}. Features such as network slicing, virtualization, and software-defined networking can be utilized to enhance the coordination of information in time-sensitive applications, such as smart transportation. Specifically for safe connected vehicles, communication with road-side network infrastructure, in the form of Vehicle-to-Infrastructure (V2I) and Infrastructure-to-Vehicle (I2V), can lead to significant improvement of safety margins through transmission of critical information about the road situation (\cite{8246845}). An interesting feature that can conveniently be utilized is virtual network slicing. Operating on the same physical hardware, virtual network slices generally consist of independent sets of software functionalities to support the service requirements for specific applications, see \cite{7926921}. These functionalities include bandwidth, speed, coverage, privacy and connectivity that can be independently implemented and optimized to satisfy particular user demands.

We propose a novel event-based add-on safety mechanism to enhance vehicle controllability at the events of occurring unexpected danger stimuli on the road. Upon occurrence of a danger, the vehicle detecting it transmits a warning signal to road-side infrastructure to be broadcasted to other vehicles that might not yet have detected the danger. Depending on how fast the communication is performed, this may result in a faster notice of the imminent danger for those vehicles that do not directly face it. Arrival of the warning signal triggers the add-on safety mechanism that, together with ADS, control the vehicle such that the danger is safely avoided with smooth braking. The communication technology at the road-side infrastructure determines the quality of information coordination service. We study two communication schemes, LTE and 5G, as the supporting network services available at base stations and evaluate their performance in safety augmentation of the vehicles. We validate our proposed scheme through extensive numerical analysis using the high-fidelity CarSim software, dedicated for vehicle simulations.

The rest of this paper is organized as follows: the model and problem scenario is described in Sec. 2. The proposed add-on safety control mechanism and the considered communication schemes are discussed in Sec. 3. Performance analyses are provided in Sec. 4, and simulation results are shown in Sec. 5. The paper is concluded in Sec. 6.

\section{System Model and Problem Scenario}

\subsection{Vehicular network model}

In a vehicular network with second-order consensus realization, the kinematic description of each vehicle $i$ yields $\dot{p}^i(t) = v^{xi}(t), \; \dot{v}^{xi}(t) = u^i(t) + \varrho^i(t)$, where $p^i, v^{xi}$ are the differences between the desired vehicle longitudinal position/speed trajectories (with a common constant speed $v_d$) and the measured/absolute ones $\bar{p}^i, \bar{v}^{xi}$ and $\varrho^i(t)$ is a zero-mean Gaussian uncertainty uncorrelated across the connected vehicle. The effect of vehicle yaw rate on the longitudinal kinematics is ignored. Thus, the measured longitudinal acceleration $a^{xi}$ can be expressed as the time derivative of longitudinal speed $\dot{\bar{v}}^{xi}$. The desired position for each vehicle is defined by $p_d^i = v_d t + s_i$, where $v_d$ is the common constant speed and $s_i$ is a given desired spacing. The vehicle speed $v_{xi}$ can be measured by GPS or estimated by using inertial measurement unit (IMU) and wheel encoders (\cite{selmanaj2017vehicle}). Inspired by the work in (\cite{oncu2014cooperative})  and \cite[Ch. 4]{tegling2018fundamental}, and having access to position and velocity errors, we define the control input for each vehicle $i$ at time $t$, with the set of neighboring vehicles denoted by $\mathcal{N}_i$, as
%-----------------------------------------------------
\begin{align}\label{eq_problem2}
u^i(t) \!=\!& -\!\!\sum_{j \in \mathcal{N}_{i}} \!\bar{c}_{ij}^p (p^i(t) \!-\! p^j(t)) \!- \!\!\sum_{j \in \mathcal{N}_{i}} \!\bar{c}_{ij}^v (v^{xi}(t) \!-\! v^{xj}(t))\\
&- c_0^{p} p^i(t) - c_0^{v} v^{xi}(t), \nonumber
\end{align}
%-----------------------------------------------------
%whereas $\rho_i$ is a random process and introduced to realize the position of emergency cases in the model. 
where $\bar{c}_{ij}^p$ and $\bar{c}_{ij}^v$ represent constant relative position and velocity gains in the conventional vehicular formation framework, and $c_{0}^p$ and $c_{0}^v$ denote the constant gains to adjust the position and velocity of vehicle $i$ w.r.t. its desired trajectories and are identical for all vehicles. 
%In order to deal with such event-based danger/emergency situations, add-on safety features is realized by additive gains $\Delta c_i^p, \Delta c_i^v$ to the position and speed gains $c_{ij}^p, c_{ij}^v$ in the conventional vehicular formation control framework. This yields $\bar{c}_{ij}^p = c_{ij}^p + \Delta c_i^p $ and $\bar{c}_{ij}^v = c_{ij}^v + \Delta c_i^v$. 
Combining the kinematics of all vehicles and the control input, the closed-loop description of the vehicular network can be written as $\dot{x} = \bar{A} x + \bar{B} \boldsymbol{\varrho}$, with the augmented state vector $x\triangleq[p,v]=[p^1,p^2,\ldots, v^{x1},v^{x2},\ldots]^{\top}$, %includes the differences between the desired longitudinal position/velocity trajectories for all vehicles 
and 
%-----------------------------------------------------
\begin{align}
    \bar{A} =
    \begin{bmatrix}
        \mathbf{0} & \mathbf{I} \\[0.3em]
        -c_{0}^p \mathbf{I} - \mathcal{L}_{c^{p}} & -c_{0}^v \mathbf{I} - \mathcal{L}_{c^v}
    \end{bmatrix}, \;
    \bar{B} = 
    \begin{bmatrix}
        \mathbf{0} \\[0.3em]
        \mathbf{I}
    \end{bmatrix},
\label{eqn:Sys_Dyn1}
\end{align}
%-----------------------------------------------------
in which $\mathcal{L}_{c^{p}}, \mathcal{L}_{c^{v}}$ are position and velocity weighted graph Laplacian matrices with weights $\bar{c}_{ij}^p$ and $\bar{c}_{ij}^v$, respectively. Since communication operates in time-slotted fashion, it is convenient to consider the discrete-time model of the described vehicular network system, which can be described as $x(k+1) = A x(k) + B \boldsymbol{\varrho}(k)$ with state and input matrices $A,B$ obtained by the zero-order hold, i.e., $A = e^{\bar{A}(t) T_s}$ and $B \!=\! \int_0^{T_s} e^{\bar{A}(t) \tau} \bar{B}(t) d\tau$, with $T_s$ denoting the sampling time. Hence, the dynamics of the vehicle at time-step $k\in \mathbb{N}\cup \{0\}$ is the sample of its continuous model at time $kT_s$, and the next sample time $k+1$ equals $kT_s+T_s$. We assume that $T_s$ is small enough, and all vehicles of the network are sampled with identical sampling times.

\subsection{Problem scenario}

\begin{figure}
    \centering
    \psfrag{a}[c][c]{\scriptsize\text{Road-side base station}}
    \psfrag{b}[c][c]{\scriptsize\text{Objects may obstruct driver's sight}}
    \psfrag{c}[c][c]{\scriptsize\text{Imminent danger}}
    \psfrag{d}[c][c]{\scriptsize\text{I2V broadcast}}
    \psfrag{ee}[c][c]{\scriptsize\text{of warning signal}}
    \psfrag{e}[c][c]{\scriptsize\text{V2I unicast}}
    \psfrag{ee}[c][c]{\scriptsize\text{of warning signal}}
    \psfrag{f}[c][c]{\scriptsize\text{Vehicle in behind}}
    \psfrag{ff}[c][c]{\scriptsize\text{may not directly detect}}
    \psfrag{fff}[c][c]{\scriptsize\text{the danger on time}}
    \psfrag{g}[c][c]{\scriptsize\text{Vehicle in front}}
    \psfrag{gg}[c][c]{\scriptsize\text{sends a warning signal}}
    \includegraphics[width=8.6cm,height=6.2cm]{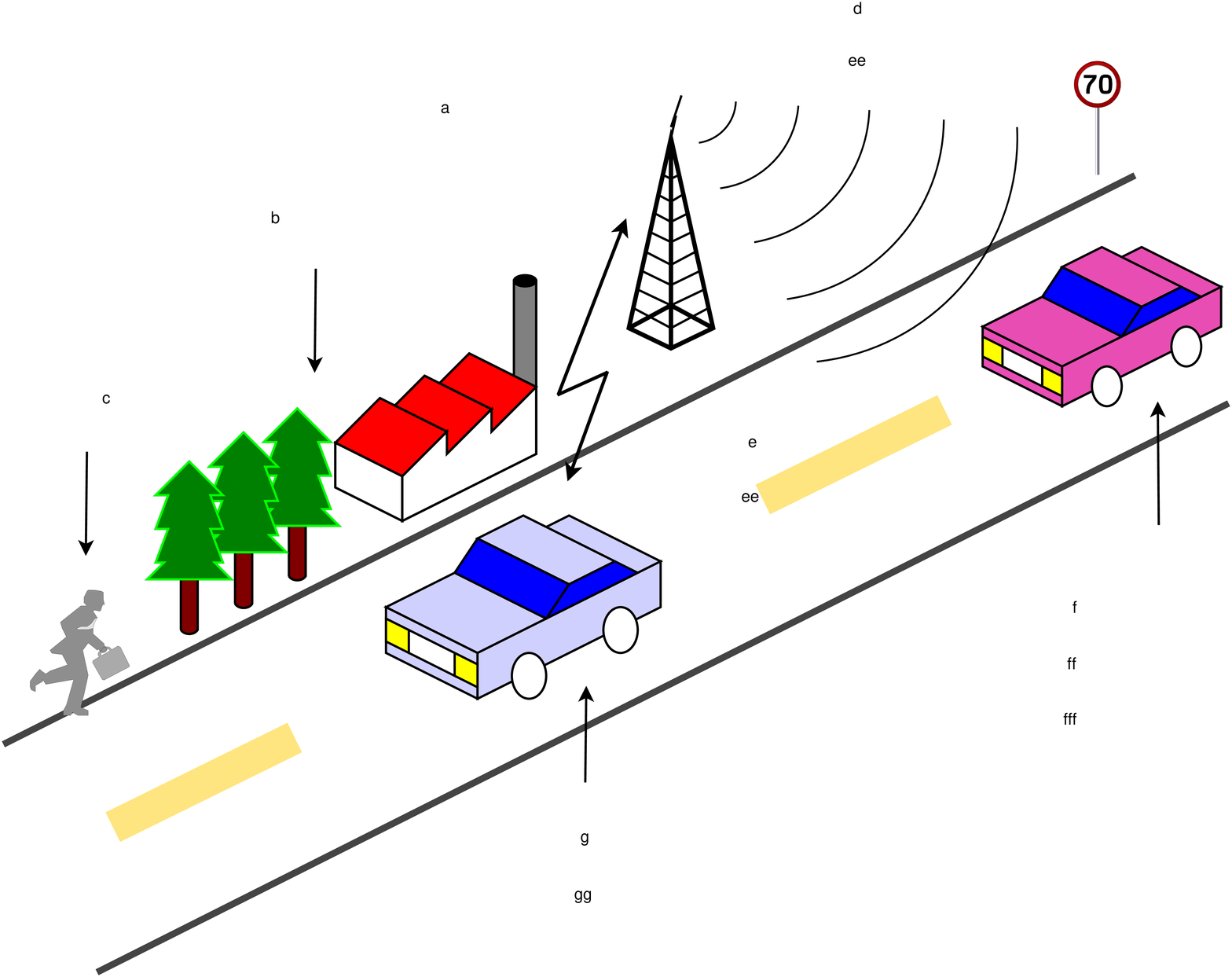}
    \vspace{-1mm}\caption{Front vehicle detects an imminent road danger which following vehicle may not detect in time. A V2I-I2V communication of a warning signal informs the following vehicles about an existing danger.}
    \label{fig:schematic}
\end{figure}

We model V2I and I2V communication via the road-side base stations to warn the neighboring vehicles about an existing danger. The considered situations are specifically the ones that, first, vehicles do not expect them and, second, the vehicles that are not directly facing the danger may not be capable of detecting them in time, e.g., if a pedestrian suddenly jumps onto the vehicle lane, a sudden congestion forms ahead, or an unusual obstacle comes into the sight (see Fig.~\ref{fig:schematic}). In these situations, the vehicles behind can be informed about the imminent danger by the vehicle facing it, so they perform appropriate actions in time to avoid harsh braking and possibly collision. The emergency information can lead to significant enhancement of braking and vehicle handling performances as it serves the vehicle (or automated driving) stability programs to generate required longitudinal forces less aggressively, which leads to less longitudinal and lateral tire force drop.

In case of confronting such poorly predictable occurrences, the vehicles that are warned about an immediate danger via communication with road-side infrastructure can improve their actions accordingly. Note that, the communicated information about an unexpected danger is an additional information exchange along with the regular V2V data exchange between the neighboring vehicles. In the rest of this paper, we refer to the unexpected danger situations and the time of their occurrences as \textit{danger} and \textit{danger time}, respectively. To incorporate the described scenario into our model, we should first notice that such on-road dangers occur sporadically with uncorrelated random frequency. Therefore, it is intuitive to assume that the probability of occurring the next danger is totally independent and uncorrelated from the previous ones.

Define by $t_n$ the time that the $n^\textsf{th}$ danger occurs. 
%Since the communication network operates in time-slotted fashion and for the ease of derivations, we model the occurrence of dangers in discrete time. Let $T_s$ be the sampling duration of vehicles' dynamics, i.e. the dynamics of the vehicle at time-step $k\in \mathbb{N}\cup \{0\}$ is the sample of its continuous model at time $kT_s$, and the next sample is taken at the time-step $k+1$ that equals $kT_s+T_s$. 
We express the sequence of danger times up to an arbitrary time-step $k$ by $\mathcal{D}_k\triangleq \{t_1,\ldots,t_n\}$, $t_n\leq k$, which ensures that the $n^\textsf{th}$ danger has been occurred up until time-step $k$. Notice that $n$ and $k$ are statistically uncorrelated. 
%The probability that the next danger occurs up until the next time-step $k+1$, can be expressed by a Bernoulli process with occurrence probability $p\in[0,1]$, as follows:
%\begin{equation}\label{eq:Bernoulli-process}
%    \textsf{P}(t_{n+1}=k+1)=p, \quad  \textsf{P}(t_{n+1}>k+1)=1-p.
%\end{equation}

For simplicity, we consider that dangers occur only in front of a vehicle and it is detectable only by the vehicle directly facing it. Each vehicle is assumed to be equipped with an add-on safety mechanism that is activated when a warning signal about an imminent danger is received from the infrastructure. In fact, the vehicle detecting a danger immediately transmits a warning signal to the infrastructure to be broadcasted to the vehicles in behind. Vehicles receiving the warning signal will activate their add-on safety mechanism and adjust their control actions accordingly to avoid collision or loss of maneuverability. We define the binary-valued triggering variable $\theta_k^i$, corresponding to the vehicle $i$ at time-step $k$, as follows
\begin{equation}\label{eq:trigger-variable}
\theta_k^i=\begin{cases} 1, & \quad\text{if} \;\;\exists \;t_n \in \mathcal{D}_k\; \text{such that:}\;\; t_n=k, \\ 0, & \quad\text{if} \;\; \forall\;  t_n \in \mathcal{D}_k: \;\; t_n<k,
\end{cases}
\end{equation}
where $\theta_k^i=1$ activates the add-on safety mechanism at time-step $k$ and it remains activated until the vehicle is at complete standstill.
%As discussed before, the described dangers occur randomly and sporadically, hence unlike the other information that each vehicle receives on a timely basis, we prefer not to monitor the danger states constantly, and instead we opt to detect, transmit, receive, and incorporate the danger states only at the time of their occurrences. To do this, we employ an \textit{event-based} danger detector to activate the danger-control mechanism as soon as a vehicle detects a danger or is informed about an imminent one.  
 Incorporating the event-triggered variable $\theta_k^i$, we introduce the additive gains $\Delta c_i^p$ and $\Delta c_i^v$ as add-on safety features to adjust the control input. The new time-varying control gains then are realized as $\bar{c}_{ij}^p(k) = \bar{c}_{ij}^p + \theta_k^i \Delta c_i^p$ and $\bar{c}_{ij}^v(k) = \bar{c}_{ij}^v + \theta_k^i \Delta c_i^v$, affecting the closed-loop behavior in $x(k+1) = A(k)x(k) + B(k) \boldsymbol{\varrho}(k)$. %We will discuss that incorporation of the event-triggered add-on safety mechanism in the vehicular consensus model can significantly improve the brake time and enhance maneuverability and as a result vehicle lateral stability due to the combined-slip tire force effect. 

The discrete-time control input $u_k^i$ for vehicle $i$ with active add-on safety mechanism, i.e. if $\theta_k^i\!=\!1$, can be stated as
\begin{align}\label{eq:controller_disc}
u_k^i =& -\!\sum_{j \in \mathcal{N}_{i}} (\bar{c}_{ij}^p+\theta_k^i\Delta c_i^p) (p_k^i - p_k^j) \\\nonumber
&- \!\sum_{j \in \mathcal{N}_{i}} (\bar{c}_{ij}^v+\theta_k^i\Delta c_i^v) (v_k^{xi} - v_k^{xj})- c_0^{p} p_k^i - c_0^{v} v_k^{xi}.
\end{align}
The augmented control input vector $u_k\triangleq[u_k^1,u_k^2,\ldots]$ of the vehicular network at time-step $k$, can be expressed as
\begin{equation}\label{eq:controller_disc_overall}
    u_k= -\begin{bmatrix}
c_0^p\mathbf{I}+\bar{\mathcal{L}}_{c^p} & c_0^v\mathbf{I}+\bar{\mathcal{L}}_{c^v}
\end{bmatrix}x,
\end{equation}
where, $\bar{\mathcal{L}}_{c^p}$ and $\bar{\mathcal{L}}_{c^v}$ are the position and velocity Laplacian matrices of the vehicular network weighted by the new gains $\bar{c}_{ij}^p+\theta_k^i\Delta c_i^p$ and $\bar{c}_{ij}^v+\theta_k^i\Delta c_i^v$, respectively. 

\begin{remark}
In line with real technology, we assume that all vehicles in the network are equipped with Anti-lock Braking Systems (ABS). The advantage of having ABS is that the slip ratio does not grow large and tire forces are kept, at maximum, around the saturation region, before which tire forces are linear. In fact, as we discuss it in Sec. 4, before the saturation region, it is convenient to consider a linear model for the slip ratio w.r.t the tire forces.
\end{remark}

%\begin{color}{blue}{
%Ehsan: Vahid jan, fortunately, I was able to activate ABS for our simulations. This make the scenario more close to real practice, since all ADAS and automated driving systems use ABS (and won't deactivate it). This provides a permission to our linear tire force assumption in the Performance Analysis section since in the worst case, it keeps the longitudinal force around the saturation region, thus, just before that point, tire forces are linear. Based on the simulation results I got, the maximum slip ratio, just before going to the nonlinear tire force region, is about 22\% at which ABS tries to keep the vehicle if the requested input is high. Please have this number in mind if you want to use it or check the results.
%%Vahid: This red-highlighted part needs to be modified. I will take care of that when the model in equation (2) is clearer for me.
%}\end{color}

%\begin{color}{blue}{
%Ehsan: Please put the following slip ratio definition somewhere in Section II if you agree:
%
%The longitudinal slip ratio $\kappa_{ir} = \frac{R_i \omega_{ir} - v_{xi}}{\max \{R_i \omega_{ir}, v_{xi}\}}$ at a corner $r$ of vehicle $i$, with $R_i$ as the effective radius of the tire, $\omega_{ir}$ as the wheel speed at the corner $r$ of vehicle $i$, and $v_{xi}$ as the longitudinal speed of (a corner of) vehicle $i$, is a good indicator of healthy brake scenarios due to its correlation with the longitudinal tire forces.
%}\end{color}

\section{Vehicle Communication Model}

In case of detecting a danger by a vehicle, a warning signal is transmitted for the in-range road side base station (BS). For technical correctness, we consider two assumptions: 
\begin{enumerate}
    \item for any vehicle there is always a road side BS in range so that the vehicle can communicate with,
    \item the task of hand over from one BS to the next is performed perfectly and instantaneously.
    %\item transmission of data from the vehicle to the BS occurs instantaneously, i.e. without any latency.
\end{enumerate}
Each BS is assumed to service a regular stationary traffic that is statistically independent from the traffic of warning signals. Warning signals are transmitted to the BS in unicast fashion, and they arrive at BS with uplink latency. 

In the following, we first introduce the scenario without V2I-I2V communication, i.e., if a danger occurs, no warning signal is transmitted through the road-side infrastructure. We then consider communication with infrastructure and discuss two wireless technologies; a non-scheduled queue-based Long Term Evolution (LTE) approach, and a resource-aware 5G network, for data coordination. We evaluate the effects of V2I-I2V communication in increasing vehicles' safety margins compared with the case that no emergency information is exchanged via infrastructure. For the purpose of comparisons, the constant gains $\bar{c}_{ij}^p$, $\bar{c}_{ij}^v$, $c_{0}^p$ and $c_{0}^v$ are assumed identical for all scenarios, and control inputs are adjusted with $\Delta c_i^p$ and $\Delta c_i^v$.

\subsection{Vehicle control without V2I-I2V communication}
Assume that in case a danger occurs on the road, no warning signal is communicated among the vehicles via the road-side infrastructure. The V2V communication, however, may exist and the following vehicle (from now on denoted by $i$) can receive the brake status of the front vehicle (from now on denoted by $j$), using the brake position sensors. For simplicity, we assume that vehicles incorporate V2V information solely from the vehicle in front.
%velocity, acceleration and  %IMU, GPS, and 
Hence, vehicle $i$ is informed about a change in brake status of vehicle $j$, i.e., if the brake of vehicle $j$ is activated. 
%the the information about vehicle $j$ available for the  includes $\{v_{xj}(k),a_{xj}(k),b_j(k)\}$, where $b_j(k)$ denotes the  at time $k$.  The brake status $b_j(k)$ informs the vehicle $i$ the time that the brake of vehicle $j$ is activated. 

\textit{Definition 1}: Driver Reaction Time (RT) is defined as the temporal duration from the time instant the driver detects a danger stimulus until an action (braking) is executed.

Characterizing RT is complicated because it is a function of many parameters such as driver's age, awareness, tiredness, eyesight, psychological state, weather and visibility condition. %Therefore, to quantify the RT, one needs to precisely take into account many parameters that are difficult to measure in real time. 
Many real and simulator-driven tests have been carried out to provide some categorically acceptable average values of RT. A main categorizations corresponds to the ``unexpectedness'' nature of the stimuli, i.e., if driver can anticipate them. It is reported in (\cite{1628-01}) that in real traffic, the mean RT for unanticipated and anticipated stimuli are $1.3 s$  and $0.7 s$, respectively. More recent results (\cite{krauss2015forensic}) indicate that for clearly-visible stimulus, about 85-95$\%$ of drivers have unanticipated RTs of about $1.5 s$. It also reports that the minimum observed RT is on average not less than $0.75 s$. In this paper, the aim is to show that even if a driver has a short RT, incorporating V2I-I2V communication can still improve vehicle safety. Moreover, we select the average RT for unexpected stimuli to be consistent with the described dangers we consider. Hence, we select a relatively short average RT of $0.7 s$ for any driver on the road. The presented results remain valid for longer RTs. 

In the absence of V2I-I2V communication, if a danger (that it cannot be directly detected by vehicle $i$) occurs in front of the vehicle $j$, then the earliest the vehicle $i$ detects it, is when the driver of vehicle $j$ reacts, i.e., after elapsing the driver's RT that vehicle $i$ receives a changed brake status of vehicle $j$. Assuming that the ADAS of vehicle $i$ reacts immediately, it takes, on average, $0.7 s$ to execute braking from the time that danger is detected by vehicle $j$. 

\subsection{Non-scheduled LTE-based V2I-I2V communication}
Now we assume that upon detecting a danger by a vehicle, it transmits a warning signal to a LTE road-side BS to be communicated with neighboring vehicles. The warning signal, when received by the ADS system of the following vehicles, immediately activates the add-on safety mechanism. We model the communication technology at the BS as a time slotted channel with limited bandwidth, i.e., at each slot, only a limited number of transmission requests can be serviced. If there are more arrivals, they are queued based on the arrival time and transmissions occur when bandwidth is assigned. Hence, some requests are delivered to the recipients by a few time slots delay. We, moreover, assume that the BS is not capable of scheduling, hence, every data arriving at the BS, either the emergency warning signals or regular traffic, are queued in a simple first-in first-out (FIFO) buffer and data is discharged depending on the queue length and the assigned bandwidth. 

We denote the arrival and departure traffics at the BS at time-step $k$ by $a_k\in \mathbb{Z}_{\geq 0}$ and $d_k \in \mathbb{Z}_{\geq 0}$, respectively. Therefore, if the queue has the length $\ell_k \in\mathbb{Z}_{\geq 0}$ at time-step $k$, the process $\ell_{k+1}$ has the following evolution
\begin{equation}\label{eq:queueing}
    \ell_{k+1}=a_k+\max[\ell_k-d_k, 0], \quad \ell_0=0.
\end{equation}

%Next, we revisit queue rate stability concept \cite[Ch. 2]{Neely2010}.

\textit{Definition 2}: A discrete time process $P(t)$ is said to be \textit{rate stable}, if 
\[\textsf{P}\left[\lim_{t\rightarrow \infty}\frac{P(t)}{t}=0\right]=1.\]

According to the \textit{Definition 2}, the process (\ref{eq:queueing}) is rate stable if and only if \cite[Th. 2.4]{Neely2010}:
\[\lim_{t\rightarrow \infty}\frac{1}{t}\sum_{\tau=0}^{t-1}[a_{\tau}-d_{\tau}]\leq 0.\]

To avoid infinite delays, we assume that queues at BSs are rate stable. Note that, rate stability is not a per-time-step condition, but a time-averaged. Although the average departure rate is greater than the average arrival rate, there might be time instances that the realized departure bandwidth cannot service all the arrived packets plus the ones already queued. Hence, data delivery would experience some delays at some finite number of instances.

Using Kendall's notation, we consider $M/M/1$ queues at the BSs, i.e., the arrival traffic at the BS follows the \textit{Poisson} distribution and the service times of the received data packets are distributed \textit{exponentially}. Compared to the previous set up, i.e. without V2I-I2V communication, here the following vehicle is informed about the existing danger subject to the uplink and downlink latency plus queuing service delay. Therefore, if after occurring a danger, the uplink/downlink latency plus service delay is shorter than the RT, the vehicle in behind receives the warning signal in shorter time by the infrastructure and can react faster to the danger. We denote the uplink delay, i.e., the time between sending the warning signal and receiving it at the BS, by $\tau_{ul}$. Similarly, we define $\tau_{dl}$ as the downlink delay, the gap between the time the warning signal is broadcasted by the BS until it is received by the vehicles. Queuing delay is also denoted by $\tau_q$, with the mean $0\leq\E[\tau_q]<1$ for pre-empty queues, see (\cite{8619752}). Therefore, if a danger is detected by vehicle $j$ at time-step $k$, the time that the transmitted warning signal is received by the vehicle $i$ in behind is $k+\tau_{ul}+\tau_q+\tau_{dl}$. For technical convenience regarding the discrete model of vehicles, the warning signal transmitted by the vehicle $j$ at time-step $k$ will be received by the vehicle $i$ at time $k+r$, $r\in \mathbb{Z}_{\geq 0}$, where $rT_s\leq \tau_{ul}+\tau_q+\tau_{dl}< (r+1)T_s$.

\subsection{5G-supported slicing-enabled V2I-I2V communication}
Here we assume that 5G-supported BSs are in range for data exchange among the vehicles. In addition, we presume that multiples of the BSs are remotely supported by a cloud data center that can provide network services such as resource management, virtualization and slicing for the local BSs. In accordance with capabilities of 5G networks, we consider a simple service slicing mechanism available for the BSs facilitating a situation-aware serviceability.\footnote{Discussing the core technology of network slicing and virtualization and their standards in 5G networks is out of scope of this paper.}

\begin{figure}
    \centering
    \psfrag{a}[c][c]{\small$a_k$}
    \psfrag{aaa}[c][c]{\small$a_k^{\text{slice 1}}$}
    \psfrag{b}[c][c]{\small$d_k$}
    \psfrag{bbb}[c][c]{\small$a_k^{\text{slice 2}}$}
    \psfrag{c}[c][c]{\scriptsize\text{(a) Non-scheduled LTE Network -- all data packets}}
        \psfrag{cc}[c][c]{\scriptsize\text{queued and discharged according to FIFO model}}
    \psfrag{d}[c][c]{\scriptsize\text{Slice 1: low-bandwidth/high-speed to service emergency traffic}}
    \psfrag{e}[c][c]{\scriptsize\text{Slice 2: high-bandwidth/normal-speed to service regular traffic}}
    \psfrag{aa}[c][c]{\small$d_k^{\;\text{slice 1}}$}
    \psfrag{bb}[c][c]{\small$d_k^{\;\text{slice 2}}$}
    \psfrag{f}[c][c]{\scriptsize\text{(b) 5G-supported slicing-enabled network}}
    \includegraphics[width=8.9cm,height=5.2cm]{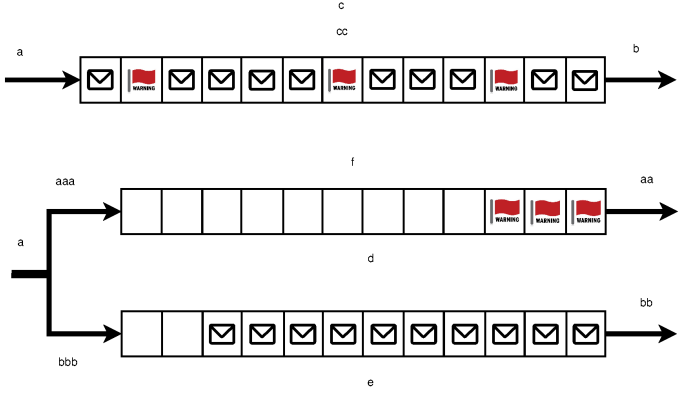}
    \caption{(a) LTE non-scheduled and (b) 5G slicing-enabled BS serviceability. Red flags and envelopes depict the warning signals and regular traffic, respectively.}
    \label{fig:communication}
\end{figure}

Employing the network slicing scheme (Fig.~\ref{fig:communication} depicts a schematic of the two discussed supporting network technologies at BS), the flagged warning signals are serviced through a specific network slice including a dedicated data transmission link. Since the described dangers occur sporadically on the road, the network slice dedicated to warning signals does not need to have a large bandwidth, however, the transmission link needs to be super fast. This entails that the communication resources (bandwidth and latency) are divided into one high-speed and low-bandwidth slice to support fast transmission of the emergency warning signals, and a regular-speed high-bandwidth slice for the transmission of regular traffic.

With the network slicing mechanism, transmission of warning signals does not suffer from queuing delay, i.e., $\tau_q=0$, though 5G networks induce a tiny amount of delay even at their fastest mode. The delay is targeted to be $1 ms$, however, reports from the operational 5G infrastructures suggest that this target has not yet been consistently achieved at the application level. For the existing physical and virtual 5G infrastructure, the latency of $5-15 ms$ is suggested for the fastest data transmission. Denote the 5G-supported latency of the fastest slice by $\tau_f$. Since uplink and downlink latency still exist, denoting $k$ as the time instance a danger is detected by vehicle $j$, the warning signal will be received by the vehicle $i$ at $k+\tau_{ul}+\tau_f+\tau_{dl}$.

\section{Performance Analysis} \label{sec:perf_analysis}

Now we investigate the advantages of the proposed infrastructure-supported add-on safety mechanism on vehicle longitudinal response in the tire levels at the event of occurring road dangers. We analyze the performance of the add-on safety features for each of the described communication scenarios on two critical parameters, \textit{longitudinal slip ratio} and \textit{stopping distance}, that essentially represent aspects of safety. The stopping distance is defined as the total traveled distance between the time that braking system is activated until the vehicle is at complete standstill. %We are interested in observing safety margin improvements for the following vehicle at the event of occurring road dangers.

Let a danger be detected by the vehicle $j$ at a time-step~$k$. The following vehicle $i$, is informed about the existing danger at time-step $k+r$, and $r$ varies depending on the selected communication scenario. For the case that no V2I communication exists (Sec. 3.1), we have $r=\lceil \frac{RT}{T_s} \rceil$. When vehicles communicate warning signals via LTE BS without scheduling (Sec. 3.2), $r$ is a function of $\tau_q$ that varies depending on the real-time queue length, arriving traffic at the BS, and the service bandwidth. Assuming constant $\tau_{ul}$ and $\tau_{dl}$, we have $r=\lceil \frac{\tau_{ul}+\tau_{q}+\tau_{dl}}{T_s} \rceil$. For the 5G-supported BS enabled with network slicing (Sec. 3.3), vehicle $i$ receives the warning signal about the danger detected by the vehicle $j$ at time-step $k$, with total latency of $r=\lceil \frac{\tau_{ul}+\tau_{f}+\tau_{dl}}{T_s} \rceil$. Denoting the time that vehicle $i$ comes at complete standstill by $\bar{k}$, the stopping distance is the distance traveled by vehicle $i$ during the interval $[k+r,\bar{k}]$.

Longitudinal slip ratio for vehicle $i$, denoted by $\kappa_k^i$, determines the tire slip in contact with road surface in longitudinal direction. As vehicles are equipped with ABS, which modulate the brake forces around the saturation region, we employ a linear model of slip ratio, as $\kappa_k^i=\frac{f_k^i}{C_i}$, where $f_k^i$ denotes the longitudinal tire forces of vehicle $i$ at time-step $k$ and $C_i$ is longitudinal tire stiffness. Assuming uniform normal load distribution and neglecting the load transfer and lateral effects of the control signal $u_k^i$, we have from the Newton's second law that $u_k^i=\frac{f_k^i}{M_i}$, with $M_i$ the effective mass of the vehicle $i$. Therefore, we obtain the slip-acceleration relation, for vehicle $i$ at time-step $k$, as
\begin{equation}\label{eq:slip_ratio}
     u_k^i=\frac{C_i}{M_i}\kappa_k^i.
\end{equation}

It is desired that the vehicle $i$ is at complete stop within a safe distance from vehicle $j$ with smooth deceleration to avoid high slip and loss of maneuverability. It should be noted that, equality (\ref{eq:slip_ratio}) is valid for low slip ratio. As ADS and vehicles are ABS-equipped, tire forces are kept around the saturation region to avoid high slip ratio, thus with this assumption, linear expression (\ref{eq:slip_ratio}) remains valid. 

Control input $u_k^i$ represents the deceleration applied to vehicle through braking, and we are interested in finding the maximum deceleration during the stopping time (the infinity norm of the sequence of control inputs during the braking process.). From the moment that the ABS is activated, it takes a short time, so called \textit{pressure build-up time} (denoted by $\tau_p$), till the maximum deceleration is reached. For the advanced braking technologies, $\tau_p$ is significantly short. For the ease of analyses, we assume that the maximum deceleration, when reached, remains constant til the vehicle comes to the complete standstill.

Assume that vehicle $i$ has the speed $v_{k+r}^i$ (we drop $x$ in the superscript for the ease of notation) when it is informed about a danger detected by vehicle~$j$. In the presence of the event-based add-on safety, $\theta_{k+r}^i\!=\!1$ and add-on safety features are activated at time $k+r$ until $\bar{k}$. Total braking time of vehicle $i$ is $\bar{k}-k-r$. As deceleration is constant after $\tau_p$, and $\theta_{t}^i\!=\!1$, $\forall t\!\in\![k+r,\bar{k}]$, the maximum deceleration of vehicle $i$ during the braking process can be obtained as
\begin{align}\label{eq:max_decel}
   |u_{\text{max}}^i|&= |u^i_t|_{\infty}=\max_{t\in[k+r,\bar{k}]}|u_t^i|\\\nonumber
     &= |-(\bar{c}_{ij}^p+\Delta c_i^p)(p_{\bar{k}-\tau_{\epsilon}}^i-p_{\bar{k}-\tau_{\epsilon}}^j)-\\\nonumber
    &\;(\bar{c}_{ij}^v+\Delta c_i^v)(v_{\bar{k}-\tau_{\epsilon}}^i-v_{\bar{k}-\tau_{\epsilon}}^j)-c_0^p\; p_{\bar{k}-\tau_{\epsilon}}^i - c_0^v \;v_{\bar{k}-\tau_{\epsilon}}^i|
\end{align}
%As the maximum deceleration is constant after the pressure build-up time, and $\theta_{k+r}^i\!=\!1$, for $t\!\in\![k+r,\bar{k}]$ we have
%\begin{align}\nonumber
%    |u^i_t|_{\infty}&=|-L_{ij}x_{k+r+\tau_p}^{ij}-L_ix_{k+r+\tau_p}^i|\\\nonumber
%    &=|-L_{ij}x_{\bar{k}-\tau_{\epsilon}}^{ij}-L_ix_{\bar{k}-\tau_{\epsilon}}^i|\\\label{eq:max_decel}
%    &= |-(\bar{c}_{ij}^p+\Delta c_i^p)(p_{\bar{k}-\tau_{\epsilon}}^i-p_{\bar{k}-\tau_{\epsilon}}^j)-\\\nonumber
%    &\;(\bar{c}_{ij}^v+\Delta c_i^v)(v_{\bar{k}-\tau_{\epsilon}}^i-v_{\bar{k}-\tau_{\epsilon}}^j)-c_0^p\; p_{\bar{k}-\tau_{\epsilon}}^i - c_0^v \;v_{\bar{k}-\tau_{\epsilon}}^i|,
%\end{align}
where, $\bar{k}-\tau_{\epsilon}$ is an infinitesimal moment before the time~$\bar{k}$ ($\tau_{\epsilon}\!\approx\! 0$), when the maximum deceleration begins to reduce. The resulting maximum slip ratio during the time interval $[k+r+\tau_p,\bar{k}-\tau_{\epsilon}]$, can be obtained from (\ref{eq:slip_ratio}) and (\ref{eq:max_decel}), as
\begin{align}
\kappa_{\text{max}}^i=\frac{M_i}{C_i} |u_{\text{max}}^i|.
\end{align}
and, $\kappa_{\text{max}}^i$ is the maximum slip determined by the ABS.

\subsection{Tuning safety parameters: desired maximum slip ratio}
The add-on safety parameters $\Delta c_i^p, \Delta c_i^v$ are generally designed based on the safe distance with front vehicle and in order to maintain smooth deceleration that leads to low slip ratio and high maneuverability of the following vehicle. To have a meaningful comparison, first we discuss safety in form of final relative distance of the vehicles when they are at complete standstill, while $\Delta c_i^p$ and $\Delta c_i^v$ are adjusted such that slip ratio dose not exceed a desired value, denoted by $\bar{\kappa}^i_{\text{max}}$. Hence, considering $\tau_{\epsilon}\approx 0$ for the ease of derivations, the following inequality is desired:
\begin{equation}\label{eq:max_s_r}
    \frac{M_i}{C_i} |u_{\text{max}}^i|\leq \bar{\kappa}^i_{\text{max}}.
\end{equation}
It is intuitive to assume that the vehicle $j$, that has detected the danger before vehicle $i$, had already come to standstill at $\bar{k}$, i.e., $v_{\bar{k}}^j=0$. Moreover, let the velocity of the vehicle $i$ at the time of being informed about the danger, i.e. $k+r$, be denoted by $v^i_{k+r}$ and assume that it remains constant till maximum deceleration is reached, i.e. $v^i_{k+r}=v^i_{k+r+\tau_p}$. Since the maximum deceleration is constant during $[k+r+\tau_p,\bar{k}]$, we can employ the quadratic equation of motion to compute the distance traveled by vehicle $i$ during the mentioned time interval, as follows:
\begin{align}\label{eq:final_pose}
p^i_{\bar{k}}&=p^i_{k+r+\tau_p}+v^i_{k+r+\tau_p}(\bar{k}-k-r-\tau_p)\\\nonumber
&+\frac{1}{2}u^i_{\text{max}}(\bar{k}-k-r-\tau_p)^2.
\end{align}
From (\ref{eq:max_decel}), (\ref{eq:max_s_r}) and (\ref{eq:final_pose}), and knowing that $v_{\bar{k}}^j=0$, we can compute the relative distance between two vehicles at stopping time for vehicle $i$, i.e., $\bar{k}$, as follows:
\begin{align}\label{eq:final_relative_distance}
    p_{\bar{k}}^i-p_{\bar{k}}^j&\leq\frac{C_i \bar{\kappa}_{\text{max}}^i}{M_i(\bar{c}_{ij}^p+\Delta c_i^p)}\left(1+\frac{(\bar{k}-k-r-\tau_p)^2}{2}\right)\\\nonumber
    &-\frac{c_0^p\left(p^i_{k+r+\tau_p}+v^i_{k+r+\tau_p}(\bar{k}-k-r-\tau_p)\right)}{\bar{c}_{ij}^p+\Delta c_i^p}.
\end{align}

It can be seen from (\ref{eq:final_relative_distance}) that, under the maximum slip ratio constraint $\bar{\kappa}_{\text{max}}^i$, the upper-bound on the distance between two vehicles at their standstill positions depends on two parameters $r$ and $\Delta c_i^p$. The first parameter is determined by the choice of the communication scenario, as discussed at the beginning of the Section 4, and $\Delta c_i^p$ is the event-triggered adjustment on the control input to augment the safety margin (relative distance). If no road-side communication exists, then we have $r=\lceil \frac{RT}{T_s} \rceil$, and since no add-on safety is considered, $\Delta c_i^p=0$ which leads to a lower value for the right side of the inequality (\ref{eq:final_relative_distance}). This means, in the absence of V2I and I2V communication, the vehicles are more prone to collide. This justifies the use of road-side infrastructure to broadcast danger information to the following vehicles. Lower $r$, that means faster communication between vehicles, increases the final relative distance. Hence, 5G with network slicing is superior to non-scheduled LTE as, first, high data rate data traffic will not affect delivery time of emergency signals in 5G, and second, queuing delay does not exist.

Increasing $\Delta c_i^p$, that essentially leads to higher deceleration, also increases the final relative distance. However, we should note that, we are not allowed to increase $\Delta c_i^p$ freely, as increasing it increases the slip ratio as well. At some point, increasing $\Delta c_i^p$ will not increase deceleration due to ABS that keeps tire forces at maximum around the saturation region. It is concluded that, road-side communication clearly enhances safety compared to the case that only V2V communication exists, while between the two described communication schemes with infrastructure, 5G generally outperforms LTE as latency is at minimum.  

Note that, the expression (\ref{eq:final_relative_distance}) does not guarantee that the vehicles never collide (i.e. $p_{\bar{k}}^i-p_{\bar{k}}^j\leq 0$) for all values of $p_{k+r+\tau_p}^i$ and $v_{k+r+\tau_p}^i$ and for any constants $C_i, M_i, \bar{c}_{ij}^p$ and $\kappa_{\text{max}}^i$. We will see this in the next section that even in the presence of add-on safety mechanism, if the initial speed of the vehicle is high and their initial distance is relatively short, then avoiding collision may not happen. However, shorter $r$ and appropriate tuning of $\Delta c_i^p$ can increase the reaction time and the maximum deceleration to make the vehicle stop in time. It is also worth mentioning that shorter $r$ results in farther back $p^i_{k+r+\tau_p}$ which increases the relative distance $p_{\bar{k}}^i-p_{\bar{k}}^j$, according to (\ref{eq:final_relative_distance}).

\begin{remark} We can repeat the above discussions for relative velocity between the two vehicles. For the purpose of brevity, we do not discuss it in detail, however, the results would be similar to that of expression (\ref{eq:final_relative_distance}), with the major exception that $\Delta c_i^v$ also appears as a tuning parameter.
\end{remark}

\subsection{Desired minimum relative distance}

Now, we consider that a given minimum relative distance between the vehicles $i$ and $j$, denoted by $p_{\bar{k}}^{i}-p_{\bar{k}}^{j}\leq p_f^{ij}$, at complete standstill is desired. Note that for a safe setup, $p_f^{ij}$ need to be negative. Having this, together with the expressions (\ref{eq:max_decel}) and (\ref{eq:final_pose}), we obtain
\begin{align}\label{eq:max_accel}
u_{\bar{k}}^i&\geq -\frac{|p_f^{ij}|\left(\bar{c}_{ij}^p+\Delta c_i^p\right)}{1+\frac{(\bar{k}-k-r-\tau_p)^2}{2}}\\\nonumber
    &+\frac{c_0^p\left(p^i_{k+r+\tau_p}+v^i_{k+r+\tau_p}(\bar{k}-k-r-\tau_p)\right)}{1+\frac{(\bar{k}-k-r-\tau_p)^2}{2}},
\end{align}
where, $|u_{\bar{k}}^i|\!\leq\!|u_{\text{max}}^i|$ should be regarded not to exceed the maximum slip ratio. According to (\ref{eq:max_accel}), for a desired $p_f^{ij}$, the maximum deceleration is a function of the tuning parameters $r$ and $\Delta c_i^p$. Similar to the Section 4.1, faster access to danger information by the following vehicle, i.e. shorter $r$, decreases the maximum required deceleration $u_{\bar{k}}^i$ to stop the two vehicles with the distance of $p_f^{ij}$. In the absence of the add-on safety mechanism, $\Delta c_i^p=0$ which results in higher deceleration. Having the event-triggered safety mechanism, we can decrease the maximum required deceleration by increasing the gain $\Delta c_i^p$ to stop the vehicles at the distance of $p_f^{ij}$. It can be deduced from (\ref{eq:max_accel}) that, the greater the desired relative distance $p_f^{ij}$ and the higher the velocity of the vehicle $i$, i.e. $v_{k+r+\tau_p}^i$, are at the time that the danger information is received, the higher the deceleration is required. Nevertheless, to guarantee maneuverability, the maximum deceleration cannot be set too high as it leads to large slip ratio. Therefore, not any given $p_f^{ij}$ is feasible. As we will show this behavior clearly in the next section, the required maximum deceleration may not be realized by the ADS system as the ABS does not allow the slip ratio to go beyond the safe region.

\section{Results and Discussions}
Sudden changes in the path planning or vehicle formation control strategies in vehicle's (and automated driving system's) dynamic stabilization programs leads to harsh brake (or throttle) actuation. As discussed in Sec. 2, high slip ratio originated by such harsh brake or acceleration requests results in considerable drop in lateral tire forces, thus makes vehicle more prone to instability. This section studies the advantages of utilizing an event-triggered add-on safety mechanism in the networked vehicular system, to reduce high-slip scenarios by proper and timely braking for two different driving cases. Validation has been done through software simulations in \textit{CarSim}, which is a system-level high-fidelity vehicle dynamics simulator. The relative position and speed signals are available through radar and vision systems in the current human-in-the-loop or automated driving systems. The ABS, which maintains tire forces around the saturation point (at the end of the linear region), is an inherent part of ADS and ADAS for networked vehicle systems and is used in simulations. This also supports the linear tire force assumption in Sec. 4.

The scenarios without V2I-I2V, non-scheduled LTE-based, and 5G-supported slicing-enabled communications, are denoted by NC, V2I, and SV2I, respectively. In simulations, vehicles are identical with mass $M \!=\! 1580$ kg, effective tire radius $R_e \!=\! 0.30$ m, and wheel base $W_b \!=\! 2.34$ m; constant and additive relative position and velocity gains are $\bar{c}_{ij}^p \!=\! 0.1, \bar{c}_{ij}^v \!=\! 0.23$ and $\Delta c_i^p \!=\! 0.04, \Delta c_i^v \!=\! 0.11$ for all vehicles; adjusting gains are $\bar{c}^p_0 \!=\! 0.08, \bar{c}^v_0 \!=\! 0.05$; the desired relative distance at standstill between two vehicles' centers of gravity is $10$~m; and the longitudinal acceleration noise is zero-mean with the variance of $0.2 \rm m/s^2$. The longitudinal slip ratio $\kappa_{l} = \frac{R_e \omega_{l} - v_{x}^a}{\max \{R_e \omega_{l}, v_{x}^a\}}$ at a (font/rear) axles $l$ of vehicle $i$, with $R_e$ as the effective radius of the tyre, $\omega_{l}$ as the average wheel speed at the axles $l$ of vehicle $i$, and $v_{x}^a$ as the longitudinal speed of vehicle $i$, is an indicator of healthy brake scenarios due to its correlation with the longitudinal tire forces; it is monitored in simulations. 

\begin{figure}[h!]
\centering
\includegraphics[width=1\linewidth]{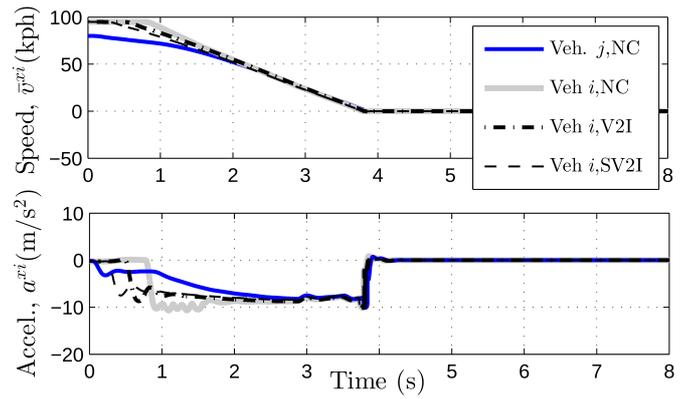}
\caption{Speed and acceleration profiles (measured) for vehicles $i$ and $j$ (blue) for NC, V2I, and SV2I scenarios}
\label{fig:VxAx_NC_V2I_SV2I}
\end{figure}

In the first case (Case 1), a longitudinal deceleration scenario, due to a danger incidence in front of the lead vehicle $j$, is simulated and the responses (slip ratio and longitudinal speed/acceleration) of the vehicle $i$ in behind are demonstrated. In Case 1, the initial speed of vehicles $j$ and $i$ are $80$ and $95$ kph, respectively. The front vehicle's ADAS activates its brakes at $t=0$, due to a danger; the reaction time for the vehicle in behind is assumed to be $0.7$ s. Then, the effect of communication is compared for non-scheduled (V2I) and 5G-supported slicing-enabled (SV2I) scenarios with $\tau_{ul} + \tau_{dl} = 0.015$ s. The queuing delay $\tau_q$ is realized from an exponential distribution with mean $0.05$ s and variance $0.015$ ms. The 5G network induces latency $\tau_f$ selected from a uniform distribution over the interval $[0.005,0.015]$ s, for the network slice that services the warning signals. The measured longitudinal speed and acceleration for vehicles $i$ and $j$ are shown in Fig.~\ref{fig:VxAx_NC_V2I_SV2I}. 
%-----------------------------------------------------
%\vspace{-10pt}
%-----------------------------------------------------

\begin{figure}[h!]
\centering
\includegraphics[width=1\linewidth]{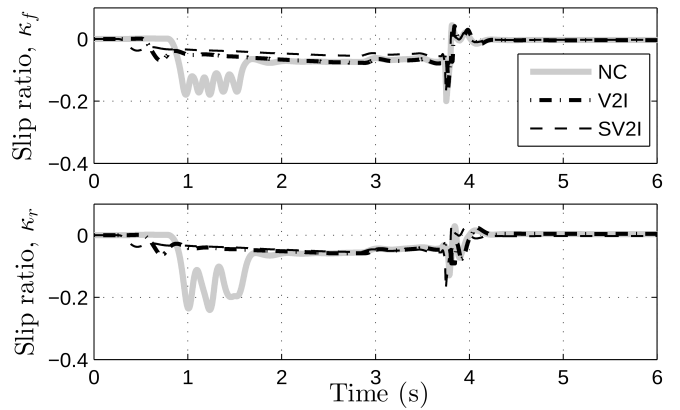}
\caption{Front and rear tires' slip ratios in Case 1}
\label{fig:kapp_NC_V2I_SV2I}
\end{figure}

As can be seen from Fig.~\ref{fig:VxAx_NC_V2I_SV2I}, front vehicle $j$ have mild deceleration, but the measured acceleration for vehicle $i$ (in behind) quickly reaches to the capacity of the dry road (almost $10 \;m/s^2$) to maintain desired spacing and stop effectively for NC scenario. This results in high slips and activation of the ABS (and consequent slip ratio oscillations) for NC scenario to modulate brake actuation around the longitudinal tire forces' saturation region (around slip ratio $22 \%$ on dry road); this is demonstrated in Fig.~\ref{fig:kapp_NC_V2I_SV2I}, in which longitudinal slip ratios for the front and rear axles of vehicle $i$ is compared for NC, V2I, and SV2I communication scenarios in Case 1. Simulation results confirm that SV2I leads to the lowest slip ratio for this driving scenario, thus facilitates better maneuverability in case of sudden cornering/lane-change request by the driver or ADS. The minor difference between the slip ratios on front and rear tires are due to load transfer during brake, which is harsher for NC, that reduces longitudinal force capacity on rear tires, thus increases rear slip ratios.    
%-----------------------------------------------------
%\vspace{-10pt}
%-----------------------------------------------------

\begin{figure}[h!]
\centering
\includegraphics[width=1\linewidth]{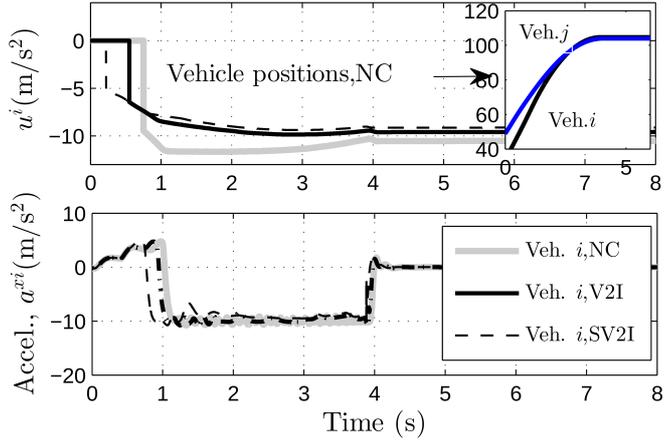}
\caption{Control inputs and measured accelerations for NC, V2I, and SV2I scenarios}
\label{fig:u_ax_NC_V2I_SV2I}
\end{figure}

\begin{figure}[h!]
\centering
\includegraphics[width=1\linewidth]{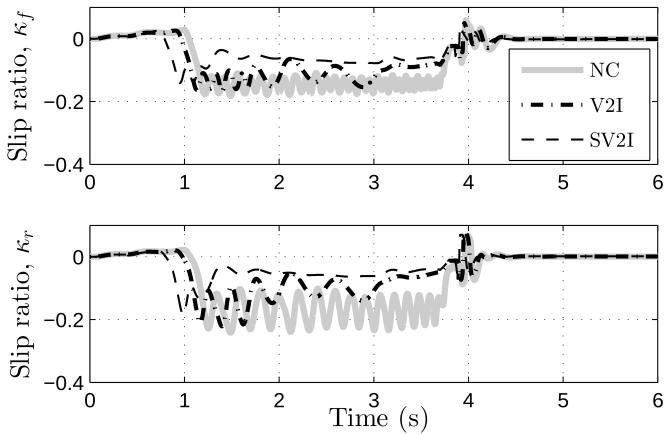}
\caption{Front/rear tyres' slip ratio for vehicle $i$ in Case 2}
\label{fig:kapp_NC_V2I_SV2I_Case2}
\end{figure}

In order to evaluate the performance of the add-on safety in harsher scenarios, in terms of maintaining the safe distance between vehicles, Case 2 is numerically simulated. In this driving case, the test setup and vehicles are similar to Case 1, but the initial speeds of vehicles $j$ and $i$ are $85$ and $110$ kph, respectively. To show the effectiveness of three scenarios in such harsh maneuver, the requested deceleration by the closed-loop system \eqref{eqn:Sys_Dyn1} and the measured one are shown in Fig.~\ref{fig:u_ax_NC_V2I_SV2I} for vehicle $i$. In Case 2 simulations, NC shows the largest deceleration request, which is not achievable due to the road-tire force capacity, thus, leads to brake modulation by the ABS around the saturation point to use the maximum attainable deceleration. However, the control system can not maintain the safe distance between two vehicles as can be seen from Fig.~\ref{fig:u_ax_NC_V2I_SV2I} (top), in which longitudinal position trajectories for both vehicles intersects at about $3.5$ s and we observe collision. Longitudinal slip for Case 2 simulations are shown in Fig.~\ref{fig:kapp_NC_V2I_SV2I_Case2}, in which NC leads to large slips during long ABS activation.

%-----------------------------------------------------
%\vspace{-10pt}
%-----------------------------------------------------
%-----------------------------------------------------
%\vspace{-10pt}
%-----------------------------------------------------
In contrast, SV2I not only maintains longitudinal slip around $7\%$, which is convenient in terms of proactive and mild brake involvement, but also keeps the safe distance between two vehicles; the test with V2I also shows lower slip ratio, when compared with NC. These results confirm effectiveness of utilizing a V2I-based event-triggered add-on safety mechanism for reducing longitudinal slip and maintaining safe distance between the vehicles.

\section{Conclusion}

In this paper, we propose an event-triggered safety mechanism that aims, together with the ADS system, to enhance the controllability of connected vehicles upon occurrence of on-road unexpected dangers. Using road-side infrastructure, danger information can be exchanged faster among the vehicles, activating the add-on safety features sooner. The control input is adjusted to avoid the danger safely, i.e. to avoid collision and brake smoothly not to lose maneuverability. We study two wireless technologies to support communication between the vehicles and the base stations, the LTE and the 5G schemes, and compare the resulting performance in terms of how fast danger information can be transmitted. We then evaluate the advantages of an add-on safety mechanism in augmenting the vehicle safety  for each communication scenario. Our results are validated through high-fidelity \textit{CarSim} software simulations.

%\begin{ack}
%Place acknowledgments here.
%\end{ack}

\bibliography{ifacconf}             % bib file to produce the bibliography
                                                     % with bibtex (preferred)
                                                   
%\begin{thebibliography}{xx}  % you can also add the bibliography by hand

%\bibitem[Able(1956)]{Abl:56}
%B.C. Able.
%\newblock Nucleic acid content of microscope.
%\newblock \emph{Nature}, 135:\penalty0 7--9, 1956.

%\bibitem[Able et~al.(1954)Able, Tagg, and Rush]{AbTaRu:54}
%B.C. Able, R.A. Tagg, and M.~Rush.
%\newblock Enzyme-catalyzed cellular transanimations.
%\newblock In A.F. Round, editor, \emph{Advances in Enzymology}, volume~2, pages
%  125--247. Academic Press, New York, 3rd edition, 1954.

%\bibitem[Keohane(1958)]{Keo:58}
%\newblock \emph{Power and Interdependence: World Politics in Transitions}.
%\newblock Little, Brown \& Co., Boston, 1958.

%\bibitem[Powers(1985)]{Pow:85}
%\newblock Is there a way out?
%\newblock \emph{Harpers}, pages 35--47, June 1985.

%\bibitem[Soukhanov(1992)]{Heritage:92}
%\newblock \emph{{The American Heritage. Dictionary of the American Language}}.
%\newblock Houghton Mifflin Company, 1992.

%\end{thebibliography}

\appendix
%\section{A summary of Latin grammar}    % Each appendix must have a short title.
%\section{Some Latin vocabulary}              % Sections and subsections are supported  
                                                                         % in the appendices.
\end{document}